# A Robust Predictive Model for Stock Price Prediction Using Deep Learning and Natural Language Processing

Sidra Mehtab
School of Computing and Analytics
NSHM Knowledge Campus
Kolkata, INDIA
email: sidramehtab.18@nshm.edu.in

Jaydip Sen
School of Computing and Analytics
NSHM Knowledge Campus
Kolkata, INDIA
email: jaydip.sen@nshm.com

*Abstract*—**Prediction of future movement of stock prices has been a subject matter of many research work. There is a gamut of literature of technical analysis of stock prices where the objective is to identify patterns in stock price movements and derive profit from it. Improving the prediction accuracy remains the single most challenge in this area of research. We propose a hybrid approach for stock price movement prediction using machine learning, deep learning, and natural language processing. We select the NIFTY 50 index values of the *National Stock Exchange* (NSE) of India, and collect its daily price movement over a period of three years (2015 – 2017). Based on the data of 2015 – 2017, we build various predictive models using machine learning, and then use those models to predict the closing value of NIFTY 50 for the period January 2018 till June 2019 with a prediction horizon of one week. For predicting the price movement patterns, we use a number of classification techniques, while for predicting the actual closing price of the stock, various regression models have been used. We also build a *Long and Short-Term Memory* (LSTM)-based deep learning network for predicting the closing price of the stocks and compare the prediction accuracies of the machine learning models with the LSTM model. We further augment the predictive model by integrating a sentiment analysis module on twitter data to correlate the *public sentiment* of stock prices with the *market sentiment*. This has been done using twitter sentiment and previous week closing values to predict stock price movement for the next week. We tested our proposed scheme using a cross validation method based on *Self Organizing Fuzzy Neural Networks* (SOFNN) and found extremely interesting results.**

## I. INTRODUCTION

Prediction of future movement of stock prices has been the subject matter of many research work. On one hand, we have proponents of the *Efficient Market Hypothesis* who claim that stock prices cannot be predicted. On the other hand, there are work that have shown that, if correctly modeled, stock prices can be predicted with a fairly reasonable degree of accuracy. The latter have focused on choice of variables, appropriate functional forms and techniques of forecasting. In this regard, Sen and Datta Chaudhauri propose a novel approach of stock price forecasting based on a time series decomposition approach of the stock prices time series [1-8].

Sen proposes a granular approach to stock price prediction in short-term time frame using machine learning- and deep learning-based models [9-10].

There is also an extent of literature on technical analysis of stock prices where the objective is to identify patterns in stock movements and derive profit from it. The literature is geared towards making money from stock price movements, and various indicators like Bollinger Band, *Moving Average Convergence Divergence* (MACD), *Relative Strength Index* (RSI), *Moving Average, Momentum Stochastics, Meta Sine Wave* etc., have been devised towards this end. There are also patterns like *Head and Shoulders, Triangle, Flag, Fibonacci Fan, Andrew's Pitchfork* etc., which are extensively used by traders for gain. These approaches provide the user with visual manifestations of the indicators which helps the ordinary investors to understand which way stock prices may move.

In this paper, we propose several machine learning and deep learning-based predictive models for predicting NIFTY 50 stock price movement in NSE of India. We use daily stock price values for three years during 2015 – 2017 for building the models, and apply the models to predict stock price movement and actual closing value of the stock with a forecast horizon of one week on the test data. For the purpose of testing, we use NIFTY 50 stock price data during the period of January 2018 till June 2019. The predictive model is further augmented by incorporating a sentiment analysis module that analyses public sentiments in Twitter on NIFTY 50 stocks. The output of the sentiment analysis module is used as the second input to the model in addition to the historical NIFTY 50 data for the purpose of predicting future stock price movement. Following the approach proposed by Mittal and Goel, we have classified the public sentiment in Twitter into four classes and studied the causal effect of those sentiment classes on NIFTY 50 stock price movement using Granger Causality Test [11].

The rest of the paper is organized as follows. In Section II, we explicitly define the problem at hand. Section III provides a brief review of the related work on stock price movement prediction. In Section IV, we describe our research methodology. Extensive results on the performance of the predictive models are presented in Section V. This describes the details of all the predictive models built in this work and the results they have produced. Finally Section VI concludes the paper.

## II. PROBLEM STATEMENT

The goal of our work is to collect the stock price of NIFTY 50 from the NSE of India and develop a robust forecasting framework for the stock price movement. We contend that price movement behavior on daily basis can be learned by powerful machine learning and deep learning-based approaches, and that knowledge can be gainfully applied for predicting future price movement of stocks. In this work, we choose the prediction horizon as one week. We hypothesize that the learning-based approaches can be further augmented by sentiment analysis of social media data in Twitter so that stock price movement can be predicted with even higher accuracy. Here, we are not addressing the problem of forecasting of short-term movement of stock price for intra-day trader. Rather, our analysis will be more relevant to a long-term investors with an objective of long-term growth of stock prices.

At any point of time in the Indian economy, given the appetite of financial market players including individuals, domestic institutions and foreign financial institutions, there is a finite amount of fund that are deployed in the stock market. This amount discounts the entire macroeconomics of that time. This fund would be distributed among various stocks. Thus, on daily basis, if some stock prices are rising, some other stock prices should be falling. Using our proposition, it will be possible for an investor to predict the movement pattern of NIFTY 50 which generally depicts the stock market sentiment in India. The approach builds in indicators like momentum, pivot points and range, all based on daily data on stock prices.

## III. RELATED WORK

The existing propositions in the literature for stock price movement and stock rice predictions can be broadly classified into three broad categories based on the choice of variables and approaches and techniques adopted in modeling. The first category includes approaches that use simple regression techniques on cross sectional data [12-14]. These models don't yield very accurate results since stock price movement is a highly non-linear process. The propositions in the second category exploit time series models and techniques like *Autoregressive Integrated Moving Average* (ARIMA), Granger Causality Test, *Autoregressive Distributed Lag* (ARDL) and *Quantile Regression* to forecast stock prices [15-17]. The third strand includes propositions using machine learning, deep learning and natural language processing for prediction of stock returns [17-19].

The drawback of the majority of the existing propositions in literature for stock price prediction is their inability to accurately predict highly dynamic and fast changing patterns in stock price movement. The current work attempts to address this shortcoming by exploiting the power of machine learning with sentiment analysis of the social media.

## IV. METHODOLOGY

In Section II, we mentioned that the goal of this work is to develop a framework for daily price movement of NIFTY 50. We collect the NIFTY 50 daily data from the period January 2, 2015 till June 28, 2019 from the Yahoo Finance website [20]. The raw data consists of the following variables: (i) *Date*, (ii) *Open* value of the index, (iii) *High* value of the index, (iv) *Low* value of the index, (v) *Close* value of the index, and (vi) *Volume* of the stock traded on a given date.

Using the six variables in the raw data, we derive the following variables that we use later on for building our predictive models. We used two approaches in forecasting - regression and classification. The two approaches involved a little differences in using some of the variables, which we will describe and explain later in this Section.

The following nine variables are derived and used in our forecasting models:

a) *month:* it refers to the month to which a given record belongs. This variable is coded into a numeric data, with "1" referring to the month of January and "12" referring to the month of December. The value of the variable *Month* lies in the range [1, 12].

b) *day_month:* this variable refers to the day of the month to which a given record belongs. It is a numeric variable lying within the range [1, 31]. For example, the date 14th February 2015 will have a value 14 against the variable *day_month*.

c) *day_week*: it is a numeric variable that refers to the day of the week corresponding to a given record. This variable lies in the range [1, 5]. Monday is coded as 1, while Friday is coded as 5.

d) *close_norm*: it is a numeric variable that is computed as a standardized value of the percentage change in the *Close* prices on two successive days. The computation of the variable is done as follows. Suppose, we have two successive days: $D_1$ and $D_2$. Let the *Close* price of the stock for $D_1$ is $X_1$ and that for $D_2$ is $X_2$. Then, *close_norm* for $D_2$ computed as $(X_2 - X_1)/X_1$ in terms of percentage.

e) *low_norm:* it is computed in the similar way as *close_norm* as the percentage change in the *Low* values over two successive days.

f) *high_norm*: it is also computed in the same way as the *close_norm* as the percentage change in the *High* values over two consecutive days.

g) *open_norm*: this variable is also computed in the same way as *close_norm* as the percentage change in the *Close* values over successive days.

h) *vol_norm*: computed as the percentage change in *Volume* over two consecutive days.

k) *range_norm*: it is computed as a percentage change in *Range* values for two successive days – the range on a day is computed as the difference between the *High* and the *Low* values on that day.

After, we compute the values of the above nine variables for each day, for the NIFTY 50 data for the period January 2, 2015 – June 28, 2019, we develop the machine learning and deep learning models for classification and regression for predictive modeling.

We use the data for the first three years (i.e., 2015 – 2017) for training the models, and test the models using the remaining data, i.e., data for the period January 2, 2018 till June 28, 2019. In regression approach, based on the historical movement of the stock prices, we predict the stock price for a horizon of one week. We use *close_norm* as the response variable, which is a continuous numeric variable. The objective of the regression technique is to predict the *close_norm* value for each day in the next week, based on the historical data of stock price movement till the current week. If the predicted *close_norm* is positive, then it will indicate that there is an expected rise the stock price on that day in comparison to its previous day, while a negative *close_norm* will indicate a fall in the stock price on the day.

In the classification approach, the response variable *close_norm* is a categorical variable having two labels. For developing the classification-based modes, we converted *close_norm* into a categorical variable that belongs to one of the two classes – "0" or "1". The value "0" indicating a negative *close_norm* value, while "1" signifying a positive *close_norm* value. Hence, if the forecast model expects a rise in the *close_norm* value on the next day, then the *close_norm* for next day will be "1". A predicted negative value of the *close_norm* on the next day will be indicated by a "0".

For both classification and regression approaches, we experimented with two cases.

**Case I:** We used the data for the period January 2, 2015 till December 29, 2017 that consisted of 737 daily records, for training the model. The predictive models are tested on the training data itself to compute accuracy of the models on the training data. Thus, in this case, we make prediction on weekly basis on the training data.

**Case II:** We apply the predictive models on the test data. The test data comprises of data for the period January 2, 2018 till June 28, 2019. The predictive power of the models is tested on the test data on prediction horizon of one week.

We use eight approaches to classification and eight approaches to regression for building our forecasting framework. The following classification models are built: (i) *Logistic Regression*, (ii) *K-Nearest Neighbor* (iii) *Decision Tree*, (iv) *Bagging*, (v) *Boosting*, (vi) *Random Forest*, (vii) *Artificial Neural Network*, and (viii) *Support Vector Machines*. The models are tested using the following metrics: (i) sensitivity, (ii) specificity, (iii) positive predictive value, (iv) negative predictive value, and (v) classification accuracy.

In similar line, we designed eight regression models using the following methods: (i) *Multivariate Regression*, (ii) *Decision Tree*, (iii) *Bagging*, (iv) *Boosting*, (v) *Random Forest*, (vi) *Artificial Neural Network*, (vii) *Support Vector Machine*, and (viii) *Long- and Short-Term Memory* (LSTM) Network. Among these approaches, LSTM is a deep learning method, while the rest are based on machine learning. For the regression methods, we use *Mean Absolute Percentage Error* (MAPE), and correlation coefficient between the actual and predicted values of the response variable (e.g., *close_norm*) as the two metrics.

For further augmenting the predictive models, we use twitter sentiment analysis of NIFTY 50 related tweets during the training and test periods. The raw tweets are first extracted from the Twitter using Twitter Streaming APIs in Python. The raw tweets are preprocessed using bag-of-words and lemmatization approaches in the SpaCy toolkit in Python. The cleaned tweets are then fed into sentiment analysis algorithm in Python that classifies the tweets into four mood classes for each day. Following the approach proposed by Mittal and Goel, the four mood classes have been defined as *calm*, *happy*, *alert* and *kind* [11]. The classified mood and the *close_norm* values of NIFTY 50 are then fed as input to the SOFNN algorithm for predicting the future values of the *close_norm* values of NIFTY 50.

Motivated by the work of Mittal and Goel, we deploy Granger Causality test in order to validate our hypothesis that mood values returned by sentiment analysis algorithm has a strong causality on the future values of the actual *close_norm* values [11]. For this purpose, we compute the *p*-values of the Granger test for several lag periods. The *p*-value here indicates the statistical significance of our hypothesis. In other words, it depicts the likelihood of getting the causality information by random chance. Thus, it is clear that lower p-values imply stronger causality and higher predictive power.

Since Granger Causality approach is based on linear models, and we suspect a high degree of nonlinearity between sentiment status measured by the mood values and the actual *close_norm* values, we focus on developing nonlinear predictive models using mood values at different lag (maximum lag value being used is 3 days) and the current value of *close_norm*. Again, following the approach of Mittal and Goel, we designed a non-linear predictive model using *Self-Organizing Fuzzy Neural Network* (SOFNN). Unlike Mittal and Goel, we use NEFCLASS for JAVA for implementing SOFNN algorithm by designing a five-layer fuzzy neural network that utilizes ellipsoidal function as its basis function [11].

For measuring the accuracy of the SOFNN-based sentiment analysis-integrated model, however we don't use the *k-fold sequential cross validation* (k-SCV) [11]. Instead, we train our predictive model on data till a given day, and test the the model by predicting the *close_norm* values for the next week, and then compute the MAPE and the correlation between the actual and predicted values of *close_norm*.

## V. PERFORMANCE RESULTS

In this Section, we provide a detailed discussion on the forecasting techniques that we have used and the results obtained using those techniques. We first discuss the classification techniques, then the regression techniques and finally the sentiment analysis based SOFNN model.

For evaluating the classification-based models, we use the following metrics:

*Sensitivity:* It is the ratio of the *true positives (correctly identified "1"s)* to the total number of *positives* in the test dataset expressed as a percentage.

*Specificity:* It is the ratio of the *true negatives (correctly identified "0"s)* to the total number of *negatives* in the test dataset expressed as a percentage.

*Positive Predictive Value (PPV):* It is the ratio of the number of *true positives* to the sum of the *true positive* cases and *false positive* cases expressed as a percentage.

*Negative Predictive Value (NPV):* It is the ratio of the number of *true negative* cases to the sum of the *true negative* cases and *false negative* cases expressed as a percentage.

*Classification Accuracy (CA):* It is the ratio of the number of cases which are correctly classified to the total number of cases expressed as a percentage.

For comparing the performance of the regression models, we use the *Mean Absolute Percentage Error* (MAPE) values and the product moment correlation values between the predicted and actual close_norm values of NIFTY 50. Tables I – VIII depict the performance results of the machine learning-based classification models.

TABLE I.    Logistic Regression Classification Results

| Stock | Case I Training Data | | Case II Test Data | |
|---|---|---|---|---|
| NIFTY 50 | Sensitivity | 42.86 | Sensitivity | 72.46 |
| | Specificity | 94.32 | Specificity | 81.08 |
| | PPV | 70.91 | PPV | 60.48 |
| | NPV | 83.62 | NPV | 88.05 |
| | CA | 81.74 | CA | 78.62 |

TABLE II.    KNN Classification Results

| Stock | Case I Training Data | | Case II Test Data | |
|---|---|---|---|---|
| NIFTY 50 | Sensitivity | 64.29 | Sensitivity | 10.63 |
| | Specificity | 94.67 | Specificity | 94.98 |
| | PPV | 79.59 | PPV | 45.83 |
| | NPV | 89.13 | NPV | 72.67 |
| | CA | 87.25 | CA | 70.90 |

TABLE III.    Decision Tree (CART) Classification Results

| Stock | Case I Training Data | | Case II Test Data | |
|---|---|---|---|---|
| NIFTY 50 | Sensitivity | 40.66 | Sensitivity | 36.23 |
| | Specificity | 92.18 | Specificity | 89.77 |
| | PPV | 62.71 | PPV | 58.59 |
| | NPV | 82.78 | NPV | 77.89 |
| | CA | 79.60 | CA | 74.48 |

TABLE IV.    Bagging Classification Results

| Stock | Case I Training Data | | Case II Test Data | |
|---|---|---|---|---|
| NIFTY 50 | Sensitivity | 68.13 | Sensitivity | 41.06 |
| | Specificity | 96.80 | Specificity | 85.33 |
| | PPV | 87.32 | PPV | 52.80 |
| | NPV | 90.38 | NPV | 78.37 |
| | CA | 89.80 | CA | 72.69 |

TABLE V.    Boosting (Adaboost) Classification Results

| Stock | Case I Training Data | | Case II Test Data | |
|---|---|---|---|---|
| NIFTY 50 | Sensitivity | 100.00 | Sensitivity | 54.60 |
| | Specificity | 100.00 | Specificity | 79.67 |
| | PPV | 100.00 | PPV | 45.89 |
| | NPV | 100.00 | NPV | 84.75 |
| | CA | 100.00 | CA | 73.66 |

TABLE VI.    Random Forest Classification Results

| Stock | Case I Training Data | | Case II Test Data | |
|---|---|---|---|---|
| NIFTY 50 | Sensitivity | 37.36 | Sensitivity | 72.95 |
| | Specificity | 88.46 | Specificity | 63.51 |
| | PPV | 51.13 | PPV | 44.41 |
| | NPV | 81.37 | NPV | 85.45 |
| | CA | 75.97 | CA | 66.21 |

TABLE VII.    ANN Classification Results

| Stock | Case I Training Data | | Case II Test Data | |
|---|---|---|---|---|
| NIFTY 50 | Sensitivity | 68.12 | Sensitivity | 7.73 |
| | Specificity | 91.31 | Specificity | 99.95 |
| | PPV | 75.81 | PPV | 99.98 |
| | NPV | 87.76 | NPV | 73.06 |
| | CA | 83.17 | CA | 73.66 |

TABLE VIII.    SVM Classification Results

| Stock | Case I Training Data | | Case II Test Data | |
|---|---|---|---|---|
| NIFTY 50 | Sensitivity | 64.71 | Sensitivity | 71.67 |
| | Specificity | 78.53 | Specificity | 75.34 |
| | PPV | 18.13 | PPV | 20.77 |
| | NPV | 96.80 | NPV | 96.72 |
| | CA | 77.58 | CA | 75.03 |

The performance results of the machine learning-based regression models are presented in Tables IX – XV.

TABLE IX.    Multivariate Regression Results

| Stock | Case I Training Data | | Case II Test Data | |
|---|---|---|---|---|
| NIFTY 50 | Correlation | 0.56 | Correlation | 0.53 |
| | MAPE | 35.02 | MAPE | 90.00 |

TABLE X.    Decision Tree Regression Results

| Stock | Case I Training Data | | Case II Test Data | |
|---|---|---|---|---|
| NIFTY 50 | Correlation | 0.98 | Correlation | 0.02 |
| | MAPE | 60.73 | MAPE | 78.37 |

TABLE XI.    Bagging Regression Results

| Stock | Case I Training Data | | Case II Test Data | |
|---|---|---|---|---|
| NIFTY 50 | Correlation | 0.70 | Correlation | 0.51 |
| | MAPE | 23.47 | MAPE | 29.32 |

TABLE XII.    Boosting Regression Results

| Stock | Case I Training Data | | Case II Test Data | |
|---|---|---|---|---|
| NIFTY 50 | Correlation | 0.69 | Correlation | 0.55 |
| | MAPE | 16.72 | MAPE | 21.34 |

TABLE XIII.    Random Forest Regression Results

| Stock | Case I Training Data | | Case II Test Data | |
|---|---|---|---|---|
| NIFTY 50 | Correlation | 0.95 | Correlation | 0.65 |
| | MAPE | 15.23 | MAPE | 19.35 |

TABLE XIV.    ANN Regression Results

| Stock | Case I Training Data | | Case II Test Data | |
|---|---|---|---|---|
| NIFTY 50 | Correlation | 0.73 | Correlation | 0.42 |
| | MAPE | 12.32 | MAPE | 25.72 |

TABLE XV.    SVM Regression Results

| Stock | Case I Training Data | | Case II Test Data | |
|---|---|---|---|---|
| NIFTY 50 | Correlation | 0.71 | Correlation | 0.75 |
| | MAPE | 17.31 | MAPE | 13.59 |

Finally, we apply a deep learning technique of regression using LSTM networks. LSTM is a variant of *Recurrent Neural Networks* (RNNs) - neural networks with feedback loops [21]. In such networks output at the current time slot depends on the current inputs as well as previous state of the network. However, LSTM overcomes the problem of vanishing and exploding gradients of RNNs during backpropagation in learning the weights of the network links [21]. We use Python programming language and the Tensorflow deep learning framework for implementing a LSTM network and utilize the network to predict the

*close_norm* values of NIFTY 50 time series on a weekly forecast horizon. We have used *Mean Absolute Error* (MAE) as the loss function and *ADAM* as the optimizer. However, we train the network with different *epoch* values and *batch sizes* of the data under different cases and for different stocks in order to obtain the optimum performance of the network.

TABLE XVI.    LSTM REGRESSION RESULTS

| Stock | Case I Training Data | | Case II Test Data | |
|---|---|---|---|---|
| NIFTY 50 | Correlation | 0.99 | Correlation | 0.99 |
| | MAPE | 8.70 | MAPE | 10.75 |

For the NIFTY 50 training data, we use 80 epochs with a batch size of 60. On completion of the epoch 72, the loss was found to have converged to as low as 0.0102, which then slowly increased to a value of 0.0678, at the end of the epoch 80. For test data set, 70 epochs with batch size 60 were found to be adequate. The loss converged to 0.0102 at the end of epoch 62, and slowly increased to 0.0251 at the completion of the epoch 70. Table XVI presents the results obtained using the LSTM-based regression model.

Now, we present the results of our sentiment analysis-augmented predictive model. First, we present the *p*-values of the Granger Causality test between the moods and the actual value of *close_norm* in the training data set at different lags. Table XVII presents the results. The mood states *calmness* and *happiness* are found to be having stronger causality on the *close_norm* values. These moods are expected to yield more accurate prediction of *close_norm*.

TABLE XVII.    GRANGER TEST p-VALUES AT DIIFERENT LAGS

| Lag | Sentiment Class | | | |
|---|---|---|---|---|
| | Clam | Happy | Alert | Kind |
| 1 | 0.0317 | 0.5867 | 0.0432 | 0.1278 |
| 2 | 0.0389 | 0.2374 | 0.2168 | 0.1325 |
| 3 | 0.0217 | 0.0874 | 0.3124 | 0.1653 |
| 4 | 0.0047 | 0.0542 | 0.3456 | 0.1732 |
| 5 | 0.0154 | 0.0673 | 0.1257 | 0.2345 |

TABLE XVIII.    PERFORMANCE OF SENTIMENT AUGMNETED ALGORITHM

| ALGO | Case I Training Data | | Case II Test Data | |
|---|---|---|---|---|
| SOFNN | Correlation | 1.00 | Correlation | 1.00 |
| | MAPE | 4.37 | MAPE | 5.37 |
| | Matched Cases | 93 % | Matched Cases | 86% |

The performance of the SOFNN algorithm with sentiment augmentation is presented in Table XVIII. First, the SOFNN model is prepared using the training data set. Then the sentiment values are provided as the second input to the model and the *close_norm* values are predicted for the next 5 days (i.e., one week). Same steps are repeated for the test dataset with same forecast horizon. The metric "matched cases" represents the percentage of cases the algorithm has been successfully able to catch the up/down pattern of the stock price (i.e., *close_norm*) movement. It may be observed that the SOFNN augmented with sentiment analysis has outperformed the deep learning-based LSTM model.

Among the classification techniques, Boosting quite expectedly, performed the best for the training data set for all metrics. However, for the test data set, Random Forest has yielded the best results, on the average. Due to imbalanced nature of the NIFTY 50 dataset, ANN is found to have performed best on specificity and PPV, while performing poorly on sensitivity. As expected, SVM has performed quite well on all metrics except PPV.

For the regression techniques, we find while LSTM outperforms all machine learning techniques, it is outperformed by the sentiment-augmented SOFNN algorithm based on natural language processing.

## VI.    CONCLUSION

In this paper, we have presented several approaches to stock price and movement prediction on a weekly forecast horizon using eight regression and eight classification methods. These models are based on machine learning and deep learning approaches. We built, fine-tuned, and tested these models using daily historical data of NIFTY 50 during January 2, 2015 till June 28, 2019. The raw data is suitably pre-processed and suitable variables are identified for building predictive models. After designing and testing the machine learning and deep learning-based models, the predictive framework is further augmented by bringing in public sentiment in the social media in addition to the historical stock prices, as the two inputs a fuzzy neural network-based SOFNN algorithm. The performance of this sentiment analysis-enhanced model is found to be the best among all models in its ability to accurately forecast the stock price movement of NIFTY 50. The study has conclusively proved that public sentiments in the social media serve as a very significant input in predictive model building for stock price movement.